\documentclass[11pt,a4paper]{article}
\usepackage{graphicx}
\usepackage[utf8]{inputenc}

\usepackage{multirow}
\usepackage{natbib}
\usepackage[noabbrev]{cleveref}

\usepackage{adjustbox}
\usepackage{array}
\usepackage{rotating}
\usepackage[flushleft]{threeparttable} 

\usepackage{tikz}
\usepackage{amssymb}
\usepackage{wasysym}

\usepackage{textcomp}

\begin{document}
\author{Niklas~Kühl, KIT, niklas.kuehl@kit.edu\\~Gerhard~Satzger, KIT, gerhard.satzger@kit.edu}
\title{Needmining: Designing Digital Support to Elicit Needs from Social Media}
\date{}
\maketitle
\begin{abstract}
Today's businesses face a high pressure to innovate in order to succeed in highly competitive markets. Successful innovations, though, typically require the identification and analysis of customer needs. While traditional, established need elicitation methods are time-proven and have demonstrated their capabilities to deliver valuable insights, they lack automation and scalability and, thus, are expensive and time-consuming.

In this article, we propose an approach to automatically identify and quantify customer needs by utilizing a novel data source: Users voluntarily and publicly expose information about themselves via social media, as for instance Facebook or Twitter. These posts may contain valuable information about the needs, wants, and demands of their authors. 

We apply a Design Science Research (DSR) methodology to add design knowledge and artifacts for the digitalization of innovation processes, in particular to provide digital support for the elicitation of customer needs. We want to investigate whether automated, speedy, and scalable need elicitation from social media is feasible. We concentrate on Twitter as a data source and on e-mobility as an application domain. In a first design cycle we conceive, implement and evaluate a method to demonstrate the feasibility of identifying those social media posts that actually express customer needs. In a second cycle, we build on this artifact to additionally quantify the need information elicited, and prove its feasibility. Third, we integrate both developed methods into an end-user software artifact and test usability in an industrial use case. Thus, we add new methods for need elicitation to the body of knowledge, and introduce concrete tooling for innovation management in practice.


\end{abstract}
\newpage

\section{Introduction}
For modern businesses to thrive,  they must design customer-centric, innovative products and services \citep{teece2010} as part of the innovation process \citep{alam2006}. To achieve this goal, providers first need to identify the needs of existing and potential future customers---so they can directly address them within their marketing activities \citep{evans1994}, and, more importantly, create innovative offerings which are tailored to those needs \citep{slater1997}.  

For \textit{need elicitation}, the application of different methods like interviews, focus groups or conjoint analyses is typical \citep{griffin1993}. These methods, however, can be very expensive and time-consuming \citep{blindheim2016}. This is mainly due to the manual effort required, as these traditional methods do neither scale well nor can they be automated.

The social change in our world, as characterized by mediatization \citep{lundby2009}, mass communication \cite{pearce2009} and participation \cite{feldmann2014} leads to a socioeconomic environment where it is common for people to publicly share personal insights on social media platforms \cite{couldry2012} like Facebook or Twitter (data availability)---and do so on an ongoing basis (data velocity). As \cite{perrin2015} shows, 65\% of all Americans and 76\% of all American internet users draw on social media networking services---with a remarkable growth occurring within the last ten years. In the group of young American adults (aged 18 to 29), already 90\% use social media. While the nature of these social media posts differs---from sharing daily updates to expressing opinions \citep{java2007}---a share contains valuable insights about the needs of customers, especially on Twitter \citep{misopoulos2014}. So fortunately for businesses, they may benefit from the fact that today customers voluntarily and intrinsically share personal information containing their authors' concerns, wants, demands, and problems. These insights are available publicly and free of cost---but any manual review would be laborious and would only allow for a snapshot at the time of the analysis. 

In this article, we deal with the research question whether customer needs can be elicited from public social media sources in an automated and scalable manner. We show the feasibility to automatically identify and quantify customer needs from the publicly available source Twitter. We develop artifacts that utilize natural language processing and predictive analytics to provide innovation managers with an aggregated overview of customer needs---and do so on an ongoing, ``always current'' basis.

In the next section, according to our DSR approach, we elaborate on the body-of-knowledge as part of the rigor cycle \citep{hevner2010}. In \Cref{sec:researchdesign}, we describe our DSR-based research design in detail, including the choices we had to make for the artifact domains. We derive the necessaty for the proposed artifact as part of an expert survey in the relevance cycle in \Cref{sec:relevance}. \Cref{sec:cycle1,sec:cycle2,sec:cycle3} are devoted to our three individual design cycles---each covering the suggestion and development as well as the evaluation of the artifact. We discuss extensions and contributions to innovation management in \Cref{discussion}, before we conclude in \Cref{sec:conclusion} with a summary, limitations and research outlook as well as theoretical and managerial implications.

\section{Rigor Cycle: Body of Knowledge}
\label{sec:relatedwork}
In order to provide common ground for the terminology of this work, to introduce related literature, and to uncover the existing research gaps, we focus on two major aspects. First, we depict the nomenclature of the terms customer needs, wants, and demands---as well as the related terms wishes, opinions, and sentiment. On that basis, we then review methods to uncover customer needs (need elicitation).

\subsection{Customer needs} 
Customer needs can be viewed from different disciplines and different perspectives, e.g. medicine \citep{doyal1991}, psychology \citep{maslow1943} or marketing \citep{oleson2004}. As this work is embedded into an economic context, the most relevant field is the latter, as it focuses on understanding the potential for adding value for the customer---as a prerequisite for successful innovation. 

\cite{kotler2001} distinguish three distinct categories of this potential: \textit{Needs} are defined as ``states of felt deprivation'', which include the most basic human requirements like food and shelter. \textit{Wants} are people's desires for specific satisfiers of needs that are shaped by societal attitudes and personal preferences. When \textit{wants} are backed by the ability to buy them and are combined with the desire for a specific product or service, they are called \textit{demands}: A person might have the \textit{need} for mobility, \textit{wants} a car, and \textit{demands} a Volkswagen Golf. From an innovation perspective, \textit{needs}, \textit{wants}, and \textit{demands} all express innovation potential with an increasing degree of specification.

Additionally, \cite{harding2001} outline that a customer need may also be latently expressed as a \textit{requirement} of a product or service—compensating  for a lack in a certain need dimension. Referring to our previous example, the person can express that he\footnote{To ensure a smooth reading flow in this work, we use only one gender and use male pronouns (he, his, him) when necessary. This always includes the female gender as well.} \textit{requires} an electric vehicle with additional rear airbags---and would preferably purchase such a vehicle if available in the market. Therefore, the person has uttered the \textit{requirement} for an airbag, while the underlying \textit{need} addressed is safety. \cite{ulwicketal2007} show which characteristics of requirements are essential to turn them into new products or services. Therefore, \textit{requirements} can easily be traced back to \textit{needs}, \textit{wants}, and \textit{demands}, and humans may reveal innovation potential in any of the four categories. While \textit{needs} are often intangible---for example, the needs for mobility or financial security-- they leave more freedom for, but also less guidance on innovation activities. Sometimes, individuals have already transformed them into wants, demands, or requirements. Thus, they give more concrete indications where innovation may be feasible, but this at the same time may conceal innovation potential: Users in the 19th century may have expressed mobility needs in terms of wants like ``faster horse-coaches'' which may have misdirected innovation efforts that led to motorized vehicles or airplanes \citep{geels2005}.

For the purpose of this work---identifying customer needs from textual (social media) data---there is little to be gained from differentiating between the different levels of needs, wants, demands, and requirements: In a first step, any information, regardless of the level of granularity, is valuable information for a marketing or innovation manager. For simplicity, we, therefore, stick with the term customer need in a broader sense---taking all four mentioned types (needs, wants, demands, and requirements) into account. 

Related to the term \textit{need} are \textit{wishes}, \textit{opinions}, and \textit{sentiments}. The New Oxford American Dictionary defines \textit{wish} as ``a desire or hope for something to happen'' \citep{mckean2005}. A wish, therefore, is an anticipated event in the future, which a person looks forward to or hopes for. Only few sources in the field of psychology try to distinguish between the term \textit{wish} and \textit{need}. \cite{shabad1993} defines a wish as something much more abstract than a need. The distinction from a need---as part of this work---is the following: A need is the desire of an individual for a certain solution to a concrete problem, whereas a wish is the desire of an individual for a certain event to take place. Nonetheless, a succinct distinction is not always possible: The method of \textit{wish mining} started off with mining wishes in terms of yearning for an event \citep{goldberg2009}, but later also considered more concrete wants, even in absence of any related event \citep{jhamtani2015}.    

An \textit{opinion}, on the other hand, typically is considered ``a view or judgment formed about something, not necessarily based on fact or knowledge'' \citep{mckean2005}. Similar definitions can be found in related literature \citep{ding2008, pang2008}. The key difference to a \textit{need} is that a judgment about something (\textit{opinion}) does not necessarily result in a desire for a concrete solution---individuals can have opinions without being affected or concerned themselves. A \textit{sentiment} is defined as a view or opinion that is held or expressed \citep{mckean2005}. Thus, \textit{sentiments} can be used to uncover and quantify opinions, as, e.g., performed in the field of \textit{opinion mining} \citep{ding2008, liu2012, cambria2013}.  

\subsection{Need elicitation}
\label{sec:needelicitation}
When it comes to identify, quantify, and prioritize customer needs as defined above, different terminologies are used. While traditional marketing sources refer to \textit{need identification} \citep{siegel1978} or \textit{need elicitation} \citep{van2005}, Design Thinking stresses the importance of \textit{needfinding} as the process of uncovering customer needs \cite{patnaik1999}. Depending on the context, this process can either mean the pure identification of needs \citep{dalgleish2014}, or also include their quantification and/or prioritization \citep{schaffhausen2015}.

Analyzing different methods of need elicitation in the literature, we can identify two distinct groups: traditional, time-proven methods as well as new, digital centric methods. In the fields of product design, service design, and marketing, traditional and well-known methods to identify and/or quantify customer needs include \textit{1-to-1 interviews} \citep{griffin1993}, \textit{lead users} \citep{hippel1986}, \textit{focus groups} \citep{topfer2008}, \textit{empathic design} \citep{leonard1997}, \textit{Kansei engineering} \citep{nagamachi1995}, \textit{house of quality} \citep{herstatt2007}, \textit{Kano model} \citep{berger1993}, \textit{persona} \citep{pirola2014}, \textit{conjoint analysis} \citep{polaine2013} as well as \textit{discrete choice} \citep{sammer2006}. More recently, though, digital-centric, analytical support for the process of need identification has emerged \citep{vanhorn2012}. Text analysis is an important method in achieving this, often applied to customer reviews \citep{hu2004}. For those reviews, a sentiment analysis allows to identify and aggregate subjective opinions (opinion mining) \citep{pang2008}. \cite{zhou2015} combine the sentiment analysis of online product reviews with a machine learning approach. While all this analytical support applies advanced digital technology, it still differs from the approach suggested in this work: It is constrained to limited product review forums where the user group is centered around a particular product, entries are known to be related to needs, and mainly incremental innovation ideas may be elicited. This is different from our approach leveraging a non product-specific platform (e.g., Twitter) that allows to holistically cover a whole domain. In addition, we target to mine from millions of tweets those``needles in the haystack'' that contain a customer need, and thus embrace all kinds of needs also allowing for radical innovation \citep{ettlie1984}. 
More related to the latter subject---finding instances with customer insights out of huge data sets and streams---is the topic of \textit{wish mining}. \cite{goldberg2009} aim at identifying the New Year's wishes for 2008 by building a ``wish detector''. Other researchers pick up this idea and apply it to customer reviews \citep{ramanand2010, wu2011, jhamtani2015}. However, these models are tailored for wishes and lack automation capability.

The definitions as used in the remainder of this work are summarized in \Cref{tab:definitions}.
\begin{table}[htbp]
	\centering
	\caption{Definitions for the remainder of this work}
	\label{tab:definitions}
	\begin{tabular}{|c|l|}
\hline
\textbf{Term}    & \multicolumn{1}{c|}{\textbf{Definition}}                                                                                                                                                                         \\ \hline
(Customer) Need  & \begin{tabular}[c]{@{}l@{}}A customer need is the desire of an individual for \\ a solution. The degree of concreteness of the desire can be \\ different, from abstract ideas to explicit demands.\end{tabular} \\ \hline
Need elicitation & \begin{tabular}[c]{@{}l@{}}Need elicitation is the process of identifying \\ and quantifying customer needs.\end{tabular}                                                                                        \\ \hline
\end{tabular}
\end{table}
\section{Research Design}
\label{sec:researchdesign}
With the necessary foundations at hand, we introduce the research design applied.  The overall Design Science Research methodology is laid out in \Cref{DSR} together with domain choices we made to build and instantiate an artifact. \Cref{rc} provides an overview of the individual design cycles covered in this work.

\subsection{Research Methodology and Application Domain}
\label{DSR}
As an overall research design, we choose Design Science Research (DSR) and base our research on \cite{hevner2010}. \citeauthor{hevner2010} suggest a DSR project should cover at least three cycles of investigation, a rigor cycle (focuses on the knowledge base, see \Cref{sec:relatedwork}), a relevance cycle (targeting the practical problem, see \Cref{sec:relevance}) and one or multiple design cycles (building and evaluating the research artifact, see \Cref{sec:cycle1,sec:cycle2,sec:cycle3}). For the individual design cycles, we follow the methodology according to \cite{kuechler2012}.

Theoretical work stresses that knowledge of customer needs is important to increase customer satisfaction when developing new products and services \citep{herrmann2000}. We draw on justificatory knowledge from the concept of \textit{fuzzy front end} \citep{khurana1997} in the innovation process, which represents the starting point where potential products and services are developed---typically on the basis of customer needs \citep{gassmann2006}. \cite{alam2006} stresses the importance of need elicitation as part of the fuzzy front end and demands more contributions in this area. As motivated previously, traditional methods of identifying customer needs tend to be expensive and time-consuming---and lack automation capabilities. We propose to address this issue by drawing on the justificatory knowledge from supervised machine learning with the deployment of a predictive model \citep{shmueli2011}. We aim to show the feasibility of using supervised machine learning with user-generated textual data as a good predictor of the existence of customer needs. 

In order to approach the task at hand, we instantiate different versions of artifacts (\textit{Needminer}) capable of the necessary preprocessing, machine learning and visualization tasks. This instantiation requires decisions on the chosen textual data source, i.e., channels where customers express needs, as well as on the application domain.

When it comes to candidate social media platforms, \cite{misopoulos2014} show the potential of Twitter for the uncovering of customer needs, which is also confirmed by \cite{kuhl2018}. With a high volume and speed of this data, e.g. 500 million tweets \citep{twitter2015} per day, Twitter presents a promising data source to gain knowledge about customer needs in order to design new products and services. Regarding the nature of these needs, \cite{marshall2015} show that people using social media tend to talk about technology-related topics. This data is a key prerequisite for instantiation of the Needminer---without the data availability the utilization of supervised machine learning would not be possible.

For testing our artifact in an application domain, we require candidate domains to be both dependent on fast and ongoing monitoring of arising needs and rich in Twitter traffic. The domain of electric mobility (e-mobility) \citep{scheurenbrand2015} fulfills both our requirements. Recent studies \citep{pfahl2013,sierzchula2014} highlight the relevancy of innovative services to foster the acceptance of e-mobility in society. However, creating innovative ideas, conceptualizing new offerings, and providing a suitable service ecosystem in that space is still a challenge \citep{hinz2015, stryja2015}. Therefore, the systematic identification of customer needs that can be addressed via new offerings is fundamental in this domain---and provides a relevant context for our work.

We further have to narrow down the domain to a geographical area with a coherent set of laws and regulations, markets as well as socio-economic conditions. In addition, we require data in a unique language, as we need consistent semantics to analyze. As a result of these requirements, languages like English and Spanish---which in general cannot be confined to one coherent region---are not suitable. Because of our familiarity with German, we focus on the German-speaking region. While there is also plenty of research on the analysis of Twitter data for the English language, there is only little research on German posts---neither on social media in general \citep{maynard2012} nor on Twitter instances in particular \citep{tumasjan2010}.

In summary, the concept of Needmining applies supervised machine learning to predict customer needs on textual data. In order to prove the feasibility of the concept, we instantiate the Needmining artifacts in this work for German tweets as the data source and for e-mobility as the application domain (c.f. \Cref{tab:positioning}).

\begin{table}[htbp] 
	\centering
	\caption{Positioning of this work}
	\begin{adjustbox}{width=\textwidth}
		{\renewcommand{\arraystretch}{1.2}
\begin{tabular}{c|l|l|l}
\cline{2-3}
\multicolumn{1}{l|}{}                                                                      & \multicolumn{2}{c|}{\textbf{Data}}                                                                                                                                                                     &                                                                                                                   \\ \cline{2-4} 
                                                                                           & \multicolumn{1}{c|}{\textbf{Source}}                                                                                                                          & \multicolumn{1}{c|}{\textbf{Language}} & \multicolumn{1}{l|}{\textbf{Domain}}                                                                              \\ \hline
\multicolumn{1}{|c|}{\begin{tabular}[c]{@{}c@{}}General\\ Needmining Concept\end{tabular}} & \begin{tabular}[c]{@{}l@{}}Any textual data source where customers\\ generate content, e.g.  Facebook, Twitter, \\ Emails, Phone Recordings etc.\end{tabular} & Any language                           & \multicolumn{1}{l|}{\begin{tabular}[c]{@{}l@{}}Any domain which is \\ discussed in digital channels\end{tabular}} \\ \hline
\multicolumn{1}{|c|}{Instantiation (this work)}                                            & Twitter                                                                                                                                                       & German                                 & \multicolumn{1}{l|}{E-Mobility}                                                                                   \\ \hline
\end{tabular}
}
	\end{adjustbox}
	\label{tab:positioning}
\end{table}

\subsection{Research Cycles}
\label{rc}
We conduct three design cycles in total. The first one addresses the question as to whether tweets containing customer needs can be automatically identified and is based on \cite{ecis2016}. This cycle is also referred to as \textit{need tweet identification}. Its purpose is to show the general feasibility of identifying tweets containing needs and it is evaluated by a \textit{technical experiment} as proposed by \cite{peffers2012}.

\begin{figure}[htbp]
	\centering
	\includegraphics[width=0.7\textwidth]{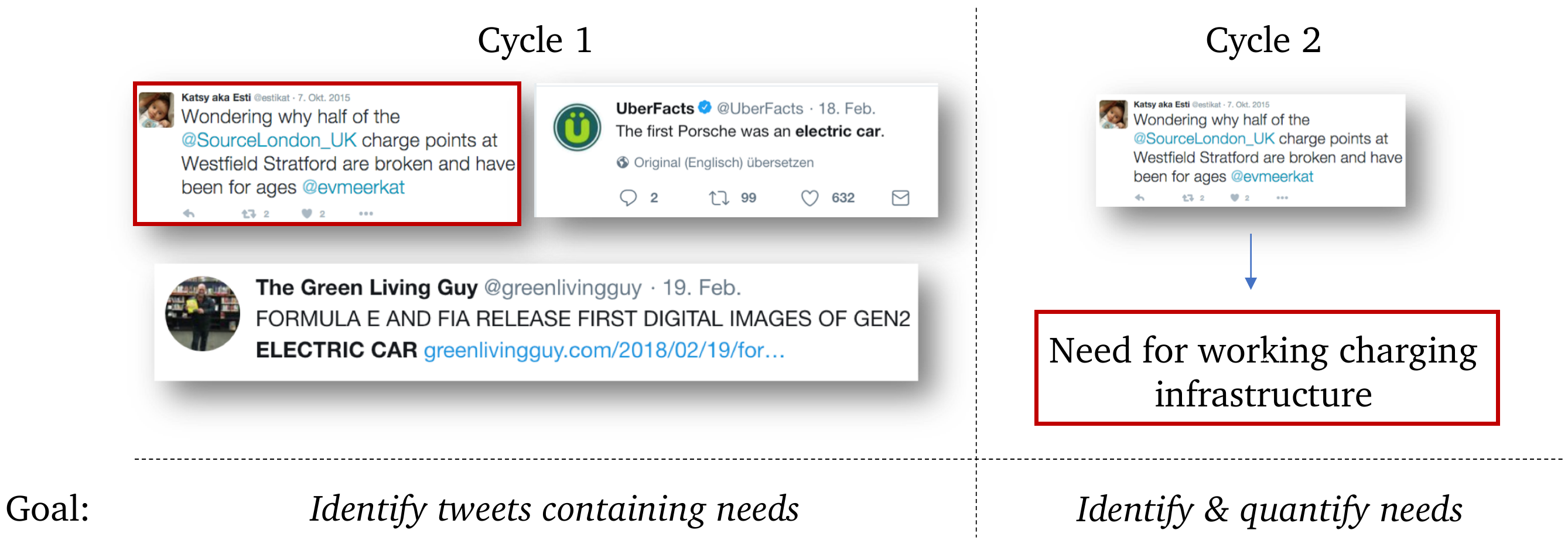}
	\caption{Goals of the first two cycles}
	\label{fig:cycles}
\end{figure}

With the feasibility shown, the second cycle addresses the allocation of tweets to needs, also referred to as the \textit{need identification and quantification}. The purpose of this cycle is to show the feasibility of extracting valuable need information from the tweets containing needs. It is based on \cite{kuhl2020supporting}. 

Finally, the third cycle (\textit{software design}) integrates the results of the first two cycles into a holistic software artifact. Suggesting and developing an end user prototype, we aim to understand the usability \citep{von2004} of the approach---and evaluate it in a \textit{workshop} with designated users in a real-world industry setting. 

\begin{figure}[htbp]
	\includegraphics[width=\textwidth]{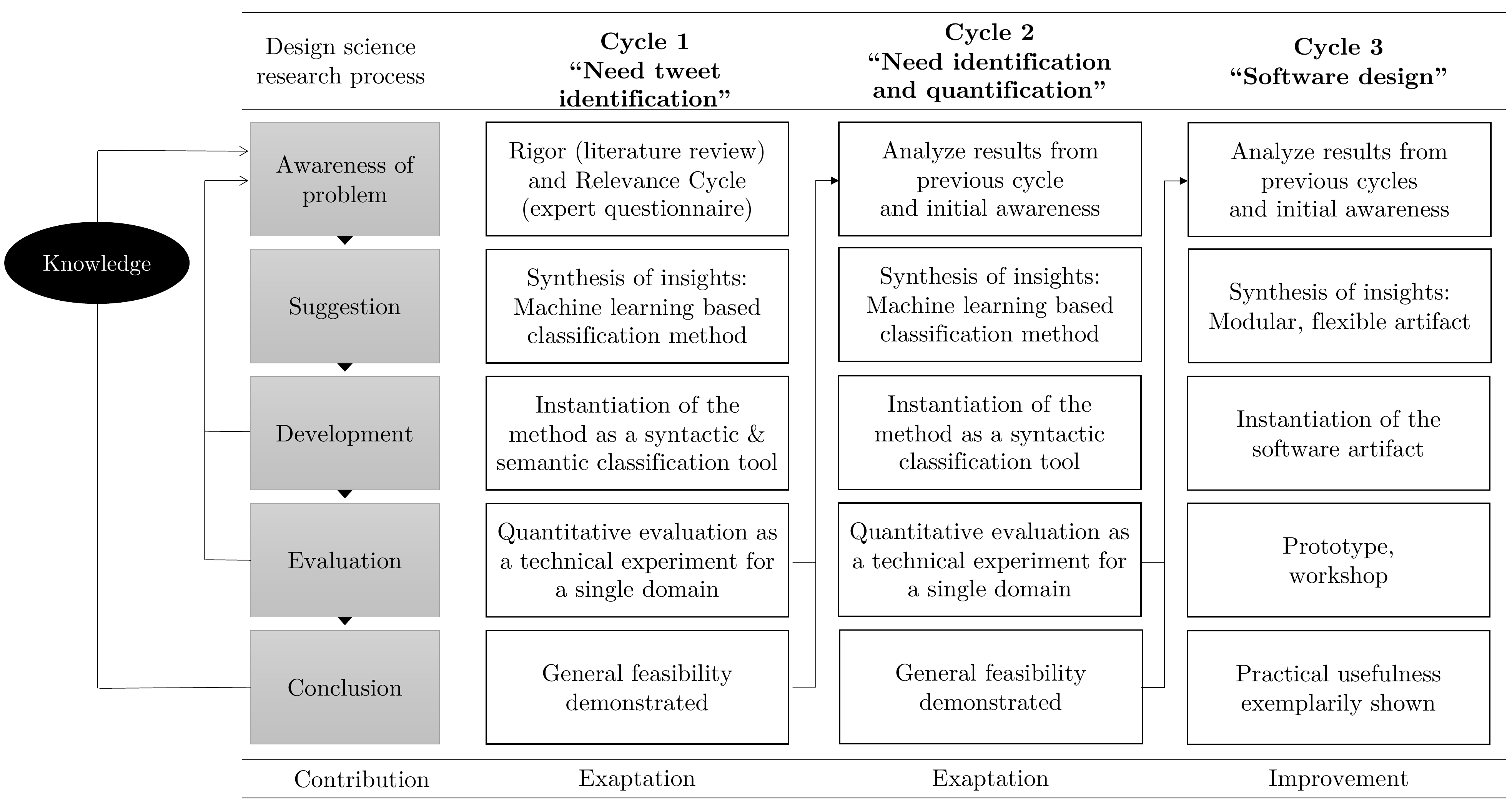}
	\caption{DSR activities of this work, based on the methodology of \cite{kuechler2012} and drawing on the evaluation types of \cite{peffers2012} and the contribution types of \cite{gregor2013}}
	\label{fig:ga-dsr}
\end{figure}

\section{Relevance Cycle: Scalable Need Elicitation}
\label{sec:relevance}

Traditionally, the elicitation of customer needs is a manual process, which can be time-consuming and cost-intensive \citep{blindheim2016}. In the context of innovation processes, such a task would be typically supervised by an innovation manager, who has a variety of methods available to elicit needs \citep{oke2007}. Furthermore, literature states possible criteria to evaluate these methods and their results. As, to the best of our knowledge, there is no comparative evaluation of these methods, we first aim to identify strengths and weaknesses of traditional methods. Also, we want to find out if automated and scalable need elicitation is of interest. We consult experts on their assessment of the different identified, traditional methods. To that end, we design an evaluation sheet for the experts as a short questionnaire \citep{glaser2010} and schedule subsequent interviews to discuss open questions or misunderstandings. As a base for a subsequent expert study, we compile an overview of existing need elicitation methods based on the taxonomy of \cite{kurniawan2004}. We then identify different evaluation criteria from literature, which form our base instrument for the consultation of experts. 

In particular, we look at the criteria of \textit{expense/effort} \citep{griffin1993}, \textit{competencies} \citep{edvardsson2012}, \textit{innovation suitability} \citep{cooper1987}, \textit{latent need detection} \citep{herstatt2007}, \textit{representativeness} \citep{bryman2015}, as well as \textit{scalability} \citep{griffin1993}. As part of our evaluation sheet, we match each method with the evaluation criteria. We use ordinally scaled ratings where applicable, as it allows for a quantitative consolidation. We choose a Likert scaling from 1 to 5 with the possibilities \textit{1 (low), 2, 3 (medium), 4, 5 (high)} \citep{bortz2007}.

We identify experts by their publications in research or practice with the aim to only identify those being familiar with all 10 methods of need elicitation. We contact 43 experts in total, of which 5 reply they are not familiar enough with the topic and 21 do not reply at all. 17 experts agree to participate and we send them the evaluation sheet and schedule interviews. Of these, 13 experts actually reply with a filled questionnaire. We then interview them individually and collect their experiences with the methods of need elicitation. After these interviews, we need to exclude 3 expert questionnaires since the corresponding experts declare they are not familiar with all mentioned methods---which is a necessity for us to compare them. Thus, we can finally draw upon questionnaires from 10 different experts.
\begin{table}[htbp]
	\caption{Results of the expert survey evaluating traditional need elicitation methods}
	\centering
	\begin{adjustbox}{width=1\textwidth}
	\includegraphics{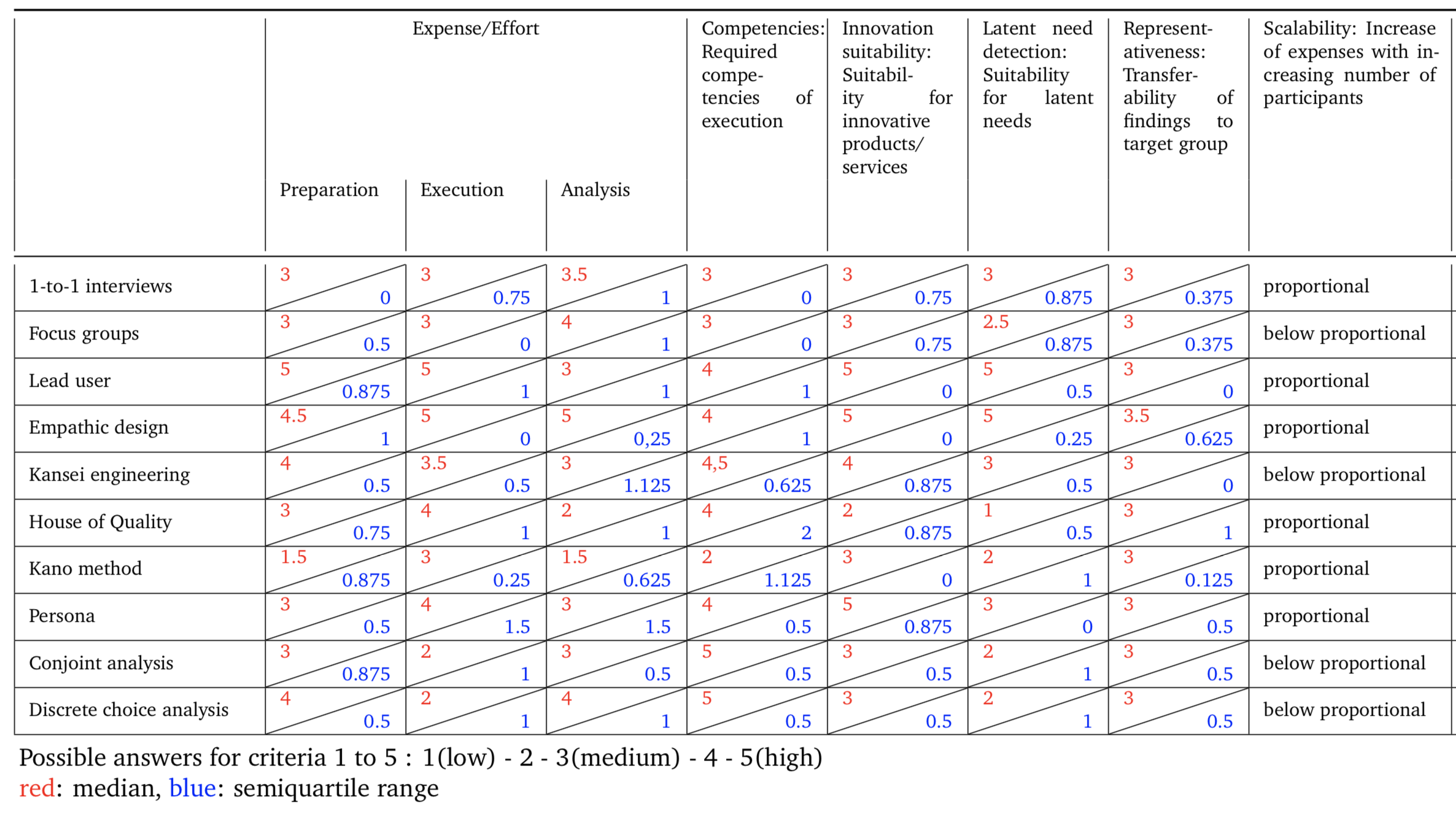}
	\end{adjustbox}
	\label{tab:needelicitation_results}
\end{table}

We come up with five particular findings from the detailed (\Cref{tab:needelicitation_results}) as well as aggregated results (\Cref{tab:aggregatedresults}): With regard to \textit{expense / effort}, any method scores at least a ``3''\footnote{For the aggregation of the results, intervals between the scales cannot be regarded as equal \citep{jamieson2004}. According to \cite{bortz2007} as well as \cite{jamieson2004}, the relevant statistical metric is the median as a measure of central tendency for ordinally scaled ratings. As a measure of statistical dispersion they, as well as \cite{raithel2006}, recommend the semiquartile range, being half the quartile distance. Apart from the quantifiable results, the nominally scaled criteria of \textit{scalability} is part of the questionnaire. When aggregating the results with regard to scalability, there is a consensus of the interviewees as all agree on the different proportionalities with one minor exception.} in minimum one of the three dimensions---meaning that there is no single method with a low effort across preparation, execution, and analysis. Thus, expense / effort is an issue with existing methods. All methods---with the notable exception of the \textit{Kano method}, score high on required competencies. Obviously, adequate support for non-experts is needed to increase diffusion of need elicitation. In terms of \textit{innovation suitability}, the median over all methods shows ``3'', with three methods being regarded as highly suitable---\textit{lead user}, \textit{empathic design}, \textit{persona}---and one being regarded as below medium (\textit{House of Quality}). In total, there are some methods more and some less suitable for innovative products and services, similar to the ability to \textit{detect latent needs}. In terms of \textit{representativeness} all methods except \textit{empathic design} are rated as medium (median of ``3'')---meaning the transferability to the target group is possible, but limited.

\begin{table}[htbp]
	\centering
	\caption{Aggregated evaluation across all regarded need elicitation methods, based on n=10 expert questionnaires}
	\label{tab:aggregatedresults}
	\begin{adjustbox}{width=\textwidth}
		\begin{tabular}{c|c|c|c|c|c|c|c|}
			\cline{2-8}
			& \multicolumn{3}{c|}{Expense/Effort} & \multirow{2}{*}{Competencies} & \multirow{2}{*}{\begin{tabular}[c]{@{}c@{}}Innovation \\ suitability\end{tabular}} & \multirow{2}{*}{\begin{tabular}[c]{@{}c@{}}Latent need \\ detection\end{tabular}} & \multirow{2}{*}{\begin{tabular}[c]{@{}c@{}}Represent-\\ ativeness\end{tabular}} \\ \cline{2-4}
			& Preparation  & Execution & Analysis &                               &                                                                                    &                                                                                   &                                                                                 \\ \hline
			\multicolumn{1}{|c|}{Median:}             & 3            & 3.25      & 3        & 4                             & 3                                                                                  & 2.75                                                                              & 3                                                                               \\ \hline
			\multicolumn{1}{|c|}{Semiquartile range:} & 0.5          & 0.5       & 0.4375   & 0.5625                        & 0.875                                                                              & 0.5                                                                               & 0                                                                               \\ \hline
		\end{tabular}
	\end{adjustbox}
	\begin{tablenotes}
		\tiny
		\item 1 (low) - 2 - 3 (medium) - 4 - 5 (high) 
	\end{tablenotes}	
\end{table}

Regarding the \textit{scalability}, no existing method is independent of the amount of participants. While some entail an underproportional increase of expense / effort with an increase of participants, some have a proportional growth. This is highly problematic, since the elicitation of customer needs from a huge amount of people ultimately results in high expenses---time- and money-wise. Therefore, we identify a lack of methods for scalable and automatable need elicitation, which cannot be covered with traditional methods.

In addition to the disadvantages of traditional need elicitation methods conducted from the questionnaire, other factors are of importance. As attitude in society, available products, services and technologies are changing more rapidly than ever \citep{ostrom2015}, the customers and their needs also change and evolve in a higher speed as they get presented with more solutions more frequently \citep{barney2001}. Moreover, these customers expect modern services to adapt faster to their needs \cite{thi2003, bughin2013}. Both factors, the high manual effort of need elicitation as well as the fast-changing universe of customer needs, result in the proposal to provide information systems which are able to automatically, and, therefore, scalable, elicit customer needs.
\section{Design Cycle 1: Need tweet identification}
\label{sec:cycle1}
As a first step of our overall approach, we evaluate the feasibility of automatic classification of Twitter data to separate out those entries that contain needs. To address this challenge, we aim to design an accurate and well-trained classification artifact, which is then able to single out need information from very large, unseen data sets in the future. By building the artifact, we gain insights on the feasibility of automatic classification of need-containing tweets. The resulting classifier then enables us to screen and elicit needs of thousands of (potential) customers---without conducting time-consuming and cost-intensive surveys or interviews.

\subsection{Suggestion and Development}
To address such automation capabilities, this first cycle evaluates the feasibility of an initial step---the identification of tweets containing customer needs (``need tweets''). We target to automatically identify these need tweets---and report on the performance. The performance is a key aspect of the classification artifact, as it determines how ``well'' it is able to differentiate between need tweets and others (precision)---and which proportion it can detect out of all need tweets (recall). We require a feasible artifact to beat simple approaches. Therfore, the baseline for our measurement is either a random guess or the simple allocation of all tweets being need tweets, the latter being the scenario someone reading through all tweets manually. To measure this performance, the pure consideration of precision or recall by themselves would not be meaningful, as an innovation manager would need a model which is both precise and also able to recall a major amount of instances---we, therefore, evaluate our model according to the $F_1$-metric \citep{vanrijsbergen1979}. The choice of performance metric could be adjusted later to fit the individual preferences of an innovation manager, for instance higher emphasis on precision ($F_{0.5}$) or recall ($F_2$).

As stated in \Cref{DSR}, we focus on German language tweets on e-mobility. Over a nineteen-month period, we collect our data set via the Twitter streaming API. We confine the domain of e-mobility via a set of filter keywords. This set comprises a list of keywords elaborated in a workshop with seven professionals in the domain (held in 2015) as well as the names of the ten most popular electric vehicles based on the registration of new cars from 2011 until 2014 in Germany \cite{neuzulassungen2014}. In total, we acquire 107,441 German\footnote{As indicated by the meta-data from Twitter: https://dev.twitter.com/streaming/overview/request-parameters, last retrieved on 2020-03-15.} tweets. We excluce bots, spam, adertisement and tweets containing URLs to receive a smaller sample for processing. We reduce the data set to 6,996 tweets for labeling. In total, 52 individuals label these filtered tweets in multiple 60-minute lab sessions with a maximum of 10 people each. Thereby, each tweet is classified three times by different participants. As a result of the labeling, we define 1,093 tweets as containing a need, requiring at least two of the participants to identify a need. We further define 4,273 tweets as not containing a need as none of the taggers identify a need. The remaining 1,630 tweets are suspended, as only one of the labelers identifies a need---and we consider this a disagreement regarding the existence of a need. An overview of the data set and its characteristics are depicted in \Cref{tab:tweetamounts}.

\begin{table}[htbp] 
	\centering
	\caption{Overview of data set}
	\begin{adjustbox}{width=1\textwidth}
		{\renewcommand{\arraystretch}{1.2}
			\begin{tabular}{|p{0.4\textwidth}|p{0.6\textwidth}|} \hline
				Keywords                             & bmw i3; e-tankstelle; eauto; ecar; egolf; electric mobility; electric vehicle; elektroauto; elektrofahrzeug; elektromobilitaet; elektromobilit{\"a}t; e-mobility; emobility; eup; fortwo electric drive; ladesaeule; lades{\"a}ule; miev; nissan leaf; opel ampera; peugeot ion; renault zoe; tesla model s \\ \hline
				Collection period                    & 2015-03-01 to 2016-05-31 \& \newline 2016-11-01 to 2017-02-28                                                                                                                                                                                                                                                                                                                                                                                                                                                                                                          \\ \hline
				Received tweets                     & 107,441                                                                                                                                                                                                                                                                                                                                                                                                                                                                                                                           \\ \hline
				Labeled tweets                 & 6,996                                                                                                                                                                                                                                                                                                                                                                                                                                                                                                                              \\ \hline
				Non-suspended tweets                    & 5,366                                                                                                                                                                                                                                                                                                                                                                                                                                                                                                                              \\ \hline
				Need tweets (share) & 1,093 (20.4\%)                                                                                                                                                                                                                                                                                                                                                                                                                                                                                                                \\ \hline                                                                                                                                                                                                               
		\end{tabular}}
	\end{adjustbox}
	\label{tab:tweetamounts}
\end{table}

We now embark on the development of a machine learning model that uses individual tweet text as input to predict the labels ``need tweet'' or ``no need tweet'' as output. 

For this, we first preprocess the data and then apply classification algorithms to it. On the basis of this data we can choose between multiple semantic and syntactic pre-processing steps \citep{maedche2001} to convert the unstructured tweet data into structured data which is digestible for supervised machine algorithms. As a methodology for supervised machine learning processes, we base our steps on \cite{hirt2017a}. An overview of the preprocessing options and their description is depicted in \Cref{tab:preprocessing}.

\begin{table}[htbp]
	\centering
	\caption{Overview of preprocessing options}
	\label{tab:preprocessing}
	\begin{adjustbox}{width=1\textwidth}
	\begin{tabular}{|l|l|l|}
		\hline
		\textbf{Type}               & \textbf{Preprocessing step}  & \textbf{Short description}                   \\ \hline
		\multirow{9}{*}{Syntactics} & Username removal             & Remove usernames, e.g. "@janboehm"           \\ \cline{2-3} 
		& Retweet tag removal          & Remove the retweet tag "RT"                  \\ \cline{2-3} 
		& Special character removal    & Remove special characters (e.g. emoticons)   \\ \cline{2-3} 
		& Minimum token length remover & Remove all tokens below a minimum length     \\ \cline{2-3} 
		& Hashtag symbol remover       & Remove the "\#" symbol                       \\ \cline{2-3} 
		& Stemming                     & Reduce words to the stem                     \\ \cline{2-3} 
		& Downcasing                   & Transform all words to lower case            \\ \cline{2-3} 
		& N-grams                      & Add contiguous sequences of n items          \\ \cline{2-3} 
		& POS Tagging                  & Tag the grammatical word classes             \\ \hline
		\multirow{4}{*}{Semantics}  & Lemmatizing                  & Reduce words to the lemma                    \\ \cline{2-3} 
		& Synset adder                 & Add the words synset(s)                      \\ \cline{2-3} 
		& Hypernym adder (level)       & Add the words hypernym(s) on a defined level \\ \cline{2-3} 
		& Disambiguator                & Choose the most probable synset for the word \\ \hline
	\end{tabular}
	\end{adjustbox}
\end{table}

As we can now obtain data sets which are transformed into a format suitable for machine learning algorithms, we can choose from different classification algorithms in systematic pre-tests. This helps us to determine a suitable candidate for a more in-depth analysis for future performance estimations. Besides a variation of preprocessing options, we choose different algorithms from the groups of Bayes classifiers, Support Vector Machines, tree-based classifiers as well as artificial neural networks \cite{michie1994}. The precise implementations are depicted in \Cref{tab:mlimplementations}. Furthermore, we can vary different sampling techniques: none, oversampling, undersampling as well as SMOTE \citep{rahman2013}.

\begin{table}[htbp]
	\centering
	\caption{Overview of implemented machine learning algorithms}
	\label{tab:mlimplementations}
	\begin{adjustbox}{width=1\textwidth}
		\begin{tabular}{|l|l|l|}
			\hline
			\textbf{Group}                                    & \textbf{Algorithm}                                                                                                                & \textbf{Implementation} \\ \hline
			\multirow{3}{*}{\textbf{Bayes classifiers}}        & Naive Bayes                                                                                                                       & \cite{john1995}       \\ \cline{2-3} 
			& \begin{tabular}[c]{@{}l@{}}Discriminative Multinomial Naive Bayes \\ (DMNB)\end{tabular}                                          & \cite{su2008}         \\ \cline{2-3} 
			& Bayes Nets                                                                                                                        & \cite{cooper1990}     \\ \hline
			\multirow{2}{*}{\textbf{\begin{tabular}[c]{@{}l@{}}Support Vector \\ Machines\end{tabular}}} & \begin{tabular}[c]{@{}l@{}}Sequential Minimal Optimization \\ based Support Vector Machines (SMO)\end{tabular}                    & \cite{platt1998}      \\ \cline{2-3} 
			& \begin{tabular}[c]{@{}l@{}}SPEGASOS: Stochastic variant of Pegasos \\ (Primal Estimated sub-GrAdient SOlver for SVM)\end{tabular} & \cite{shalev2007}     \\ \hline
			\multirow{2}{*}{\textbf{Tree-based classifiers}}  & Random Trees                                                                                                                      & \cite{breiman2001}    \\ \cline{2-3} 
			& Random Forests                                                                                                                    & \cite{breiman2001}    \\ \hline
			\multicolumn{2}{|l|}{\textbf{Artificial neural network}}                                                                                                                              & \cite{villa2012}      \\ \hline
		\end{tabular}		
	\end{adjustbox}
\end{table}

For the pre-study, we experiment with different combinations of preprocessing, sampling and algorithms---all evaluated with a five-fold cross-validation. At a mean $F_1$-score of $0.5042$, the best-performing algorithm is a Random Forest applying preprocessing with both syntactic and semantic steps. All syntactic preprocessing steps prove helpful except POS tagging, bigrams and trigrams. With regard to semantic preprocessing, synset adding and disambiguation yield the most promising results. While we cannot cover all options here, we have covered a broad set of preprocessing, sampling and algorithm options to allow for a well-chosen method to be subject to the following feasibility evaluation.

\subsection{Evaluation}
For the model performance estimation, we analyze how the performance results of our model vary with different algorithm parameters. To do so, we conduct a grid search in combination with a nested cross validation to tune parameters \citep{beleites2008}. As we iterate through $450$ parameter combinations on the inner iterations and validate the best-performing parameters on the outer folds, we achieve an estimation of the model performance as well as its robustness. The results are depicted in \Cref{tab:reserror}.
\begin{table}[htbp]
	\centering
	\caption{Performance estimation results from nested cross-validation}
	\label{tab:reserror}
	\begin{adjustbox}{width=1\textwidth}
		\begin{tabular}{|l|l|l|l|l|l|l|l|l|l|}
			\hline
			\multicolumn{1}{|c|}{\multirow{2}{*}{\textbf{\begin{tabular}[c]{@{}c@{}}\#Outer\\ fold\end{tabular}}}} & \multicolumn{4}{c|}{\textbf{Random Forest parameters}}                                                                                                                           & \multicolumn{5}{c|}{\textbf{Statistical performance}}                                                                                                                                          \\ \cline{2-10} 
			\multicolumn{1}{|c|}{}                                                                                 & \multicolumn{1}{c|}{\textbf{p (bag size)}} & \multicolumn{1}{c|}{\textbf{l (\#trees)}} & \multicolumn{1}{c|}{\textbf{K (\#features)}} & \multicolumn{1}{c|}{\textbf{d (tree depth)}} & \multicolumn{1}{c|}{\textbf{Accuracy}} & \multicolumn{1}{c|}{\textbf{AUC}} & \multicolumn{1}{c|}{\textbf{Precision}} & \multicolumn{1}{c|}{\textbf{Recall}} & \multicolumn{1}{c|}{\textbf{$F_1$-score}} \\ \hline
			1                                                                                                      & 0.8                                        & 400                                       & 100                                          & 200                                      & 0.746                                  & 0.792                             & 0.405                                   & 0.664                                & 0.503                            \\ \hline
			2                                                                                                      & 0.6                                        & 200                                       & 100                                          & 500                                      & 0.759                                  & 0.826                             & 0.425                                   & 0.692                                & 0.527                            \\ \hline
			3                                                                                                      & 0.8                                        & 100                                       & 100                                          & 500                                      & 0.751                                  & 0.792                             & 0.412                                   & 0.673                                & 0.510                            \\ \hline
			4                                                                                                      & 1                                          & 500                                       & 100                                          & 400                                      & 0.782                                  & 0.835                             & 0.459                                   & 0.697                                & 0.553                            \\ \hline
			5                                                                                                      & 0.6                                        & 100                                       & 100                                          & 100                                      & 0.766                                  & 0.806                             & 0.428                                   & 0.633                                & 0.511                            \\ \hline
		\end{tabular}
	\end{adjustbox}
\end{table}

We are able to achieve a mean $F_1$-score of $0.521$ on the 5 outer folds. The minimum score is $0.503$ and the maximum $0.553$, the standard deviation is $0.020$. These results show that the model has very low tendency towards over-fitting, since the results do not differ significantly. To prove the feasibility of our first cycle, we benchmark this result against three different baselines, which are depicted in \Cref{tab:baseline}. The first baseline is a random guess classifier, which \textit{has knowledge} of the need distribution ($20\%$ need tweets, $80\%$ no-need tweets). On average, this classifier achieves an $F_1$-score of $0.2$. With our best-performing model, we are able to show an increase of $160.5\%$. Another random guess classifier, which has \textit{no knowledge} about the distribution of need tweets, is able to achieve an average $F_1$-score of $0.286$, guessing with a probability of $0.5$. The improvement of our classifier compared to this specific guessing classifier is $82.17\%$. Lastly, we could simply assign all tweets as need tweets as a baseline. In reality, this would mean someone (e.g. the innovation manager) would need to read all tweets, knowing 80\% of them do not even contain a need. A different performance metric, taking into account the costs of reading these false positives, could be implemented in the future to further penalize such classifiers. Nonetheless, this simple assignment results in an average $F_1$-score of $0.333$, which is the best result to benchmark against. Our classifier registers an improvement of $56.30\%$ in comparison. Therefore, the feasibility of the need tweet identification is shown, as we reach statistically superior predictions of need tweets on unseen data.

\begin{table}[htbp]
	\centering
	\caption{Comparison of different baselines to results of this work}
	\label{tab:baseline}
	\begin{adjustbox}{width=1\textwidth}
		\begin{tabular}{|l|l|l|l|}
			\hline
			\textbf{Classifier}     & \textbf{Description}                                                                                                               & \textbf{$F_1$-score} & \textbf{Improvement} \\ \hline
			Random guess (p=0.2)    & \begin{tabular}[c]{@{}l@{}}Classifier with knowledge about\\  the distribution of need tweets \\ (probability=0.2)\end{tabular}    & 0.2                      & +160.50\%            \\ \hline
			Random guess (p=0.5)    & \begin{tabular}[c]{@{}l@{}}Classifier without knowledge about \\ the distribution of need tweets\\  (probability=0.5)\end{tabular} & 0.286                    & +82.17\%             \\ \hline
			Simple assignment (p=1) & \begin{tabular}[c]{@{}l@{}}Classifier which assigns all instances \\ as a need tweet\\ (probability=1)\end{tabular}                & 0.333                    & +56.30\%             \\ \hline
			This work               & Random forest classifier                                                                                                           & 0.521                    & -                    \\ \hline
		\end{tabular}
	\end{adjustbox}
\end{table}
\section{Design Cycle 2: Need identification and quantification}
\label{sec:cycle2}
After the application of a suited classifier to (continuously incoming) Twitter data, we consequently receive only\footnote{This, of course, is depending on the classifier performance. For the next cycle, we assume we are able to further develop the need tweet detection up to a perfect level.} sets of need tweets out of the universe of all tweets. This remaining set contains all relevant need information, but the tweets would still need to be read and interpreted in a manual fashion. In order to ``extract'' the needs out of the need tweets and to quantify them in order to subsequently present them in a holistic software artifact, this chapter deals with a machine-learning based approach to tackle the challenge of automatically identifying and quantifying needs from need tweets.

\subsection{Awareness of Problem}
While the previous cycle shows the feasibility of identifying whether a tweet contains a need, the need itself remains undetected. Aiming for an information system capable of automatically identifying and quantifying customer needs from Twitter data, it is crucial to not only identify the pure existence of a customer need---but more precisely be able to display the expressed needs in an aggregated fashion. To do so, we differentiate two scenarios. In this article, we regard the first scenario of quantification where the needs are already given (\textit{quantification of already known needs}). It deals with analyzing the occurrence of needs among the population, e.g. to monitor need development over time. A different scenario would be an automated identification of previously unknown needs themselves, which is, however, not part of this work due to its high complexity involving semantics and clustering.

\subsection{Suggestion and Development}
We tackle the challenge of \textit{quantifying already known needs}---and do so with a supervised machine learning approach. We continue to work with our exemplary data set from the previous cycle, in particular the 1,093 need tweets (see \Cref{tab:tweetamounts} on page \pageref{tab:tweetamounts}). Based on \cite{saldana2015}, we start with the labeling of the need tweets to identify broader concepts by descriptive coding with independent researchers. After six iterations of coding, we reach saturation and reveal 28 different needs that can be grouped into 8 major need categories:
\begin{itemize}
	\item \textit{price}: car price, electrical price, oil/gas price, price (other)
	\item \textit{car characteristics}:  car performance, driving experience, car sound, car smell, car comfort, car design, car characteristics (other)
	\item \textit{charging infrastructure}: charging infrastructure (general), charging infrastructure existence, charging infrastructure availability (physical), charging infrastructure availability (technical)
	\item \textit{range}: range
	\item \textit{charging technology}: charging interfaces and technologies, range extender, battery (other),charging speed
	\item \textit{environment \& health}: environmentally friendly car production, environmentally friendly car usage, environment \& health (other)
	\item \textit{society}: politics, unspecified desire for e-mobility
	\item \textit{other}: joke, other (miscellaneous)
\end{itemize}

We decide to continue to train our machine learning algorithm on the eight major need categories only, as there is not sufficient data available for each particular need. This circumstance also determines the granularity of our automatic need detection for the final software artifact.

Based on the procedure of the first cycle, we experiment with different preprocessing, sampling and algorithm strategies for each need category individually. After the pre-tests, we choose a Support Vector Machine for the performance evaluation and perform a grid search as part of a nested cross-validation for each of the eight needs. In total, we optimize eight different models, one for each need.

\subsection{Evaluation}
To demonstrate the feasibility of (multiple) classifiers for the assignment of need tweets to need categories, we evaluate the performances of the individual classifiers after a nested cross-validation. We benchmark these performances against the best-performing random guess, which is---as in the first cycle---a random classifier which assigns all instances the positive class, being the equivalent of a person manually reading through the need tweets and categorizing them.

The ten outer folds of this nested cross-validation result in ten different $F_{1}$-scores of different classifiers on unseen data. \Cref{tab:results_nested} shows the minimum, maximum, mean and standard deviation of these ten scores for each need category. Furthermore, we show the scores achieved by the best performing baseline classifier (``best baseline''). Depending on the need, the classifier is able to improve the classification by up to $+379.65\%$. All classifiers perform significantly better than the best baseline and, therefore, the feasibility is shown.

\begin{table}[htbp]
	\centering
	\caption{$F_1$-scores and improvements of the nested CV for the model error estimation}
	\label{tab:results_nested}
	\begin{adjustbox}{width=1\textwidth}
		\begin{tabular}{|l|r|r|r|r||l|l|} \hline
			\textbf{Need category}                                                & \multicolumn{1}{l|}{\textbf{Min}} & \multicolumn{1}{l|}{\textbf{Max}} & \multicolumn{1}{l|}{\textbf{Mean}} & \multicolumn{1}{l||}{\textbf{\begin{tabular}[c]{@{}l@{}}Standard\\ Deviation\end{tabular}}} & \textbf{\begin{tabular}[c]{@{}l@{}}Best\\ baseline\end{tabular}} & \textbf{\begin{tabular}[c]{@{}l@{}}Impro-\\ vement\end{tabular}} \\ \hline
			price                                                        & 0.524                             & 0.737                             & 0.642                              & 0.059                                                                                      & 0.304                                                         & +111.18\%                                                                             \\ \hline
			\begin{tabular}[c]{@{}l@{}}car\\ characteristics\end{tabular}        & 0.308                             & 0.600                             & 0.471                              & 0.089                                                                                      & 0.221                                                         & +113.12\%                                                                             \\ \hline
			\begin{tabular}[c]{@{}l@{}}charging\\ infrastructure\end{tabular}   & 0.719                             & 0.871                             & 0.783                              & 0.043                                                                                      & 0.429                                                         & +82.52\%                                                                             \\ \hline
			range                                                        & 0.538                             & 0.917                             & 0.721                              & 0.122                                                                                      & 0.221                                                         & +226.24\%                                                                             \\ \hline
			\begin{tabular}[c]{@{}l@{}}charging\\ technology\end{tabular}   & 0.222                             & 0.444                             & 0.360                              & 0.078                                                                                      & 0.190                                                         & +89.47\%                                                                             \\ \hline
			\begin{tabular}[c]{@{}l@{}}environment\\ \& health\end{tabular} & 0.125                             & 0.800                             & 0.543                              & 0.203                                                                                      & 0.113                                                         & +379.65\%                                                                             \\ \hline
			society                                                      & 0.452                             & 0.746                             & 0.543                              & 0.080                                                                                      & 0.400                                                         & +35.50\%                                                                              \\ \hline
			other                                                        & 0.171                             & 0.457                             & 0.278                              & 0.079                                                                                      & 0.182                                                         & +52.75\%  \\ \hline                                                                           
		\end{tabular}
	\end{adjustbox}
\end{table}
The scores for \textit{charging infrastructure, range, price} and \textit{society} show that it is possible to automatically allocate major need categories from tweets. Being the harmonic mean of precision and recall, a high $F_{1}$-score means that the classifiers are in most cases able to correctly identify whether a tweet belongs to a need category or not. This could certainly add substantial value to companies and help innovation managers in their decision making. 

The results differ notably between the individual need categories. One influencing factor is that the class distributions for each need category vary: For the need category \textit{environment \& health} for example, only 71 need tweets are available for the training, validation and testing procedure. The low score for the category \textit{others} is consistent with the fact that this category is of very diverse content---making it difficult for the classifier to find the right patterns which determine whether a tweet belongs to this category or not. 

However, in total we are able to show that the automatic quantification of needs is well feasible. Future work needs to further explore unsupervised approaches, which do not require any training in advance and are able to automatically identify the need categories.

\section{Design Cycle 3: Software Design}
\label{sec:cycle3}
Within the overall research effort of automatically identifying customer needs from Twitter data, the first DSR cycle evaluates the feasibility of need tweet identification as a basis. The second DSR cycle requires this classifier to output need tweets only, as we analyze possibilities on how to identify the needs themselves from this output. As a result of the second cycle, we achieve an additional set of classifiers which is able to automatically assign tweets to multiple need categories---which shows the feasibility of need tweet quantification.

In this last cycle, we now integrate all models into a comprehensive Needmining software artifact. With this resulting artifact, we aim to evaluate usability.

\subsection{Awareness of Problem}
The classifiers from the previous cycles are only tailored to IT-focused users, since all other activities would need to be undertaken manually. These include the automatic collection of tweets, their storage, the orchestration of the classifiers as well as any visualization tasks. \cite{mason2014} point out the importance of centralized solutions for users to design successful applications. According to \cite{endsley2016}, a central and clear user interface plays an important role.

In order to explore the capabilities of the Needmining approach, we suggest a software artifact which is able to combine all required functionalities within one service.

\subsection{Suggestion and Development}
\begin{figure}[htbp] 
	\centering
	\includegraphics[width=1\textwidth]{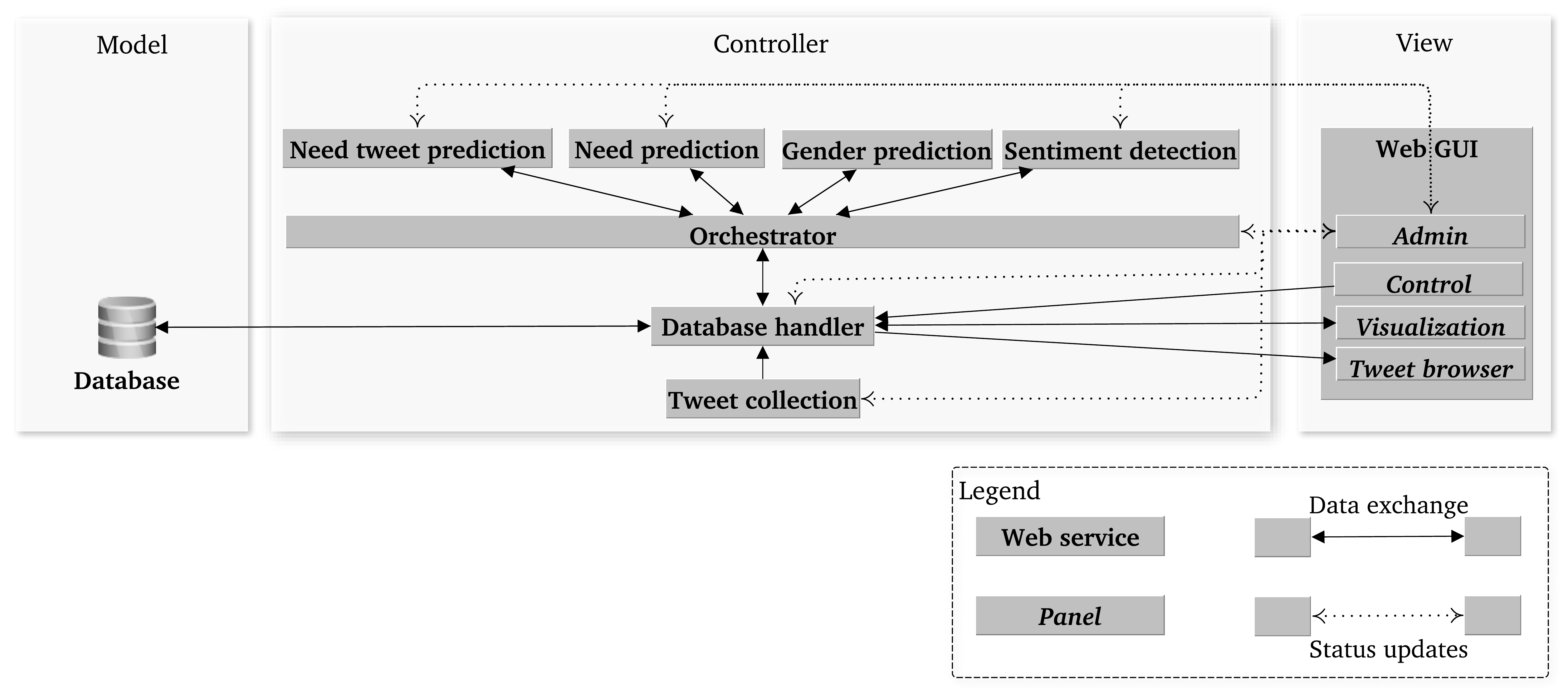}
	\caption{Needmining artifact architecture}
	\label{3-architecture}
\end{figure}

The Needmining artifact aims at automatically visualizing the most frequently expressed customer needs, which are autonomously posted on Twitter. For a general architecture of the software design, we follow two principles: model-view-controller (MVC) \citep{kransner1988} and service-oriented architecture (SOA) \citep{huhns2005, ralyte2015}. \Cref{3-architecture} depicts the general overview of the suggested architecture and its different web services. We structure our general architecture into three layers (model, view, controller) and a number of web services as independent instances, which can be accessed via a well-defined interface.

\begin{figure}[htbp] 
	\centering
	\includegraphics[width=\textwidth]{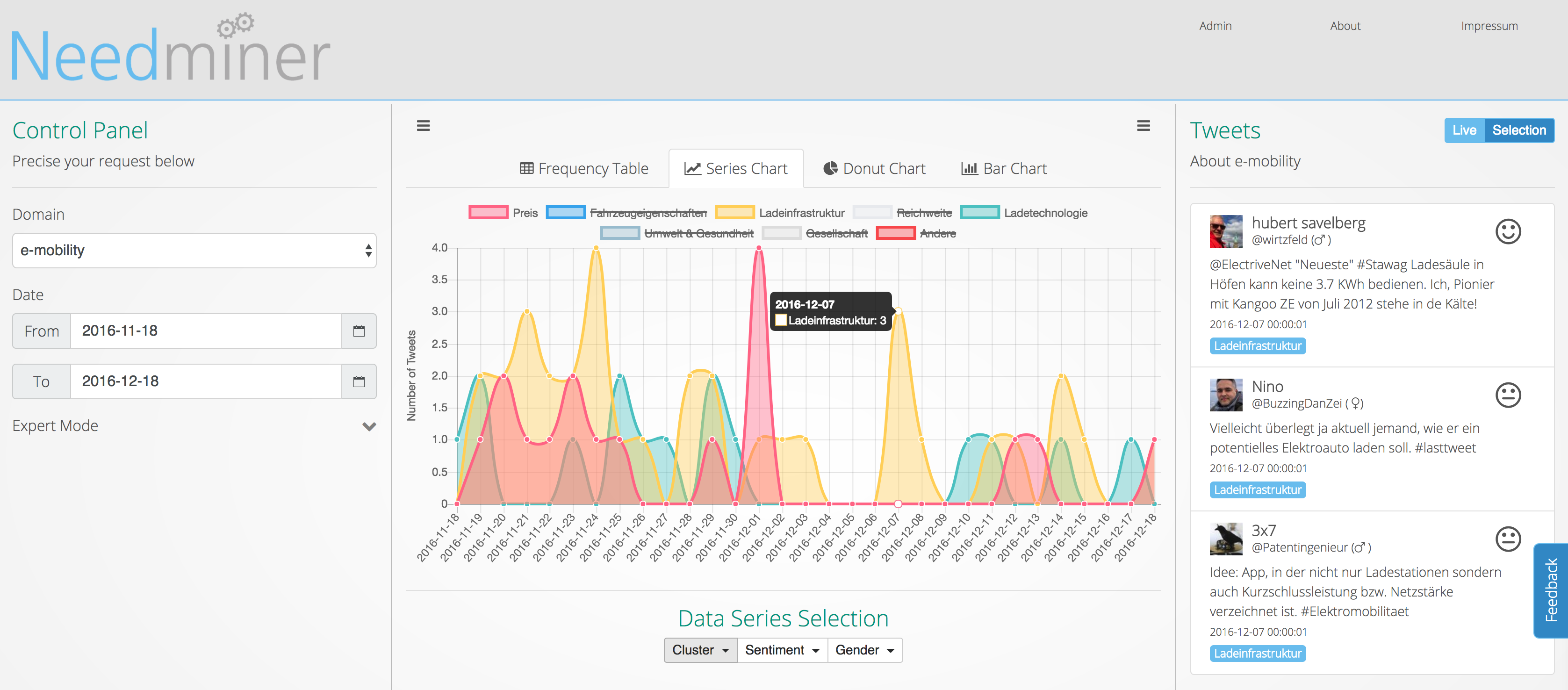}
	\caption{Screenshot of the GUI}
	\label{fig:3-needminer}
\end{figure}

These web services include, in addition to the two classifiers from the first two cycles, an automatic tweet collection, a database handler, an orchestrator, a service for predicting the gender of the tweet's author \citep{hirt2017b}, a sentiment detection \citep{thelwall2013} as well as a GUI. We design and implement the services accordingly. A screenshot of the final GUI is depicted in \Cref{fig:3-needminer}.

\subsection{Evaluation}
\label{3-evaluation}
As it is important to evaluate the artifact in a real-world scenario, we conduct an industry workshop with a large German utility and e-mobility services provider. The participants are from different divisions with different roles in the company. In total, seven employees take part in the three hour workshop. The group consists of a product \& innovation manager, a senior project manager, the head of a competence center, a group leader, a user researcher and an R\&D project manager as well as an intern.

To evaluate the prototype, all participants can use the Needmining artifact for up to 30 minutes. Then, they fill out a further questionnaire regarding the experiences with using the software and we openly discuss positive and negative feedback. The questionnaire contains questions to assess
\begin{itemize}
	\item the helpfulness of the software,
	\item its possible application within the company,
	\item its ability to replace existing methods as well as
	\item feedback regarding functionalities and design.
\end{itemize}
Out of seven participants, six consider the Needmining artifact as a helpful software artifact. Furthermore, five participants can imagine to use it for their daily business within their company. None of the participants states that the Needmining artifact in its current implementation can fully replace existing need elicitation methods.
\section{Discussion}
\label{discussion}
With the feasibility of the approach being demonstrated across the three cycles, we now discuss two topics which are of relevance for the practical use and the application potential. First, we discuss the generalization of Needmining by conucting a brief robustness check on a second application domain and evaluating the domain-specificity of the training data used. Second, we critically assess the possible enhancement of Needmining to the ``toolbox'' of need elicitation methods.

\subsection{Robustness check: Applicability to other domains}
The feasibility of classifying tweets as to whether they contain customer needs is a major contribution of the first cycle of this work. Nonetheless, the classification is only tested in the domain of e-mobility and, thus, might not be applicable to other domains without significant setup / calibration effort. The generalization of Needmining, including the domain independence of a classifier, is of major importance. Therefore, this discussion contributes to the idea of a domain-independent classification by introducing an additional domain to enable cross-domain comparisons. We analyze the domain-dependence of the ``need tweet'' classification models, i.e. analyze how specific the language for the expression of needs in a particular domain is, based on the example of the two domains \textit{e-mobility} and \textit{rail traffic}. While evaluating how well a trained model from one domain performs on the other, we do not gain an ultimate degree of generalization, but uncover first insightful views on how people express needs across different domains.

\begin{table}[htbp]
	\centering
	\caption{General intra- and cross-domain possibilities in training and testing}
	\label{tab:intra}
	\begin{adjustbox}{width=0.6\textwidth}
		
		\begin{tabular}{lc|c|c|}
			\cline{3-4}
			& \multicolumn{1}{l|}{} & \multicolumn{2}{c|}{\textbf{Training in}}                                                                                                   \\ \cline{2-4} 
			\multicolumn{1}{l|}{}                                                                                & \textbf{Domain}       & \textbf{e-mobility}                                                 & \textbf{rail traffic}                                                 \\ \hline
			\multicolumn{1}{|c|}{\multirow{2}{*}{\textbf{\begin{tabular}[c]{@{}c@{}}Testing\\ on\end{tabular}}}} & \textbf{e-mobility}   & \begin{tabular}[c]{@{}c@{}}intra-domain\\ (e-mobility)\end{tabular} & cross-domain                                                          \\ \cline{2-4} 
			\multicolumn{1}{|c|}{}                                                                               & \textbf{rail traffic} & cross-domain                                                        & \begin{tabular}[c]{@{}c@{}}intra-domain\\ (rail traffic)\end{tabular} \\ \hline
		\end{tabular}
		
	\end{adjustbox}
\end{table}

We now use various scenarios as depicted in \Cref{tab:intra}. First, we train on a reduced\footnote{We reduced the number of tweets to match the total number of tweets collected for the rail traffic scenario.} dataset from the e-mobility domain and test on a different domain and vice versa (cross-domain). Second, we train and test with data from the same domain (intra-domain). Additionally, we also combine both data sets and analyze the results. By performing such an analysis, we gain more information on how individuals express their needs across different domains on Twitter---and obtain insights on whether or not domain-independent classification is feasible. A similar approach was previously applied by \cite{goldberg2009} who analyzes wishes across different domains.

Analogous to the domain of e-mobility, we collect tweets on the topic of \textit{rail traffic} on the basis of selected keywords in June 2016. 2,396 tweets are labeled on whether they contain a customer need. We define 172 tweets as containing a need (at least two labelers identify a need) and 1,967 as not containing a need. We apply identical preprocessing and machine learning techniques as in the first cycle and depict the results of a Random Forest classifier for each scenario in \Cref{tab:intraresults}.

\begin{table}[htbp]
	\centering
	\caption{Results regarding domain independency}
	\label{tab:intraresults}
	\begin{adjustbox}{width=0.7\textwidth}
		
	\begin{tabular}{ccc|c|c||c|}
		\cline{4-6}
		&                                                                                                        &              & \multicolumn{2}{c||}{\textbf{Training data}}            & \multirow{2}{*}{\textbf{F1-score}} \\ \cline{4-5}
		& \textbf{}                                                                                              &              & \textbf{Total}         & \textbf{Need tweets (share)}  &                                    \\ \hline
		\multicolumn{1}{|c|}{\multirow{4}{*}{\textbf{Approach}}} & \multicolumn{1}{c|}{\multirow{3}{*}{\textbf{\begin{tabular}[c]{@{}c@{}}intra-\\ domain\end{tabular}}}} & e-mobility   & 2,396                  & 332 (13.9\%)                  & 0.485                              \\ \cline{3-6} 
		\multicolumn{1}{|c|}{}                                   & \multicolumn{1}{c|}{}                                                                                  & rail traffic & 2,396                  & 172 (7.2\%)                   & 0.422                              \\ \cline{3-6} 
		\multicolumn{1}{|c|}{}                                   & \multicolumn{1}{c|}{}                                                                                  & combination  & \multirow{2}{*}{4,792} & \multirow{2}{*}{504 (10.5\%)} & 0.428                              \\ \cline{2-3} \cline{6-6} 
		\multicolumn{1}{|c|}{}                                   & \multicolumn{2}{c|}{\textbf{cross-domain}}                                                                            &                        &                               & 0.284                              \\ \hline
	\end{tabular}
		
	\end{adjustbox}
\end{table}

The intra-domain results show similar performances for both domains in terms of $F_1$-score, although the score in rail traffic is slightly lower (0.422 vs. 0.485). This indicates that our e-mobility results can be replicated in other domains as well. The combination of both domains results in better performances in comparison to rail traffic, but less performance in comparison to e-mobility. This result is not surprising, as it originates from a mixture of both data sets.

The cross-domain validation evaluates the classification model’s ability to capture and apply cross-domain relevant features for instances from unknown domains. According to our measurements, we achieve a maximum $F_1$-score of $0.284$. This still denotes an improvement of $+50.3 \%$ over a baseline, which results from averaging the two baselines of the individual domains. Generally speaking, the results show a strong loss in classification performance compared to need identification within one domain. However, there must be domain-comprehensive need expression features that can be identified without domain-specific language. Future work needs to identify those features that boost a domain-independent classification---and include the models in an updated artifact. As of now, however, we would recommend building domain-specific classification models as their performance results are more convincing.

\subsection{Potential of Needmining as an enhancement of need elicitation methods}

Finally, we assess the potential of Needmining by consulting the evaluation criteria for need elicitation methods as introduced in \Cref{sec:relevance} on page \pageref{sec:relevance}.

\begin{figure}[htbp]
	\centering
	\includegraphics[width=0.8\textwidth]{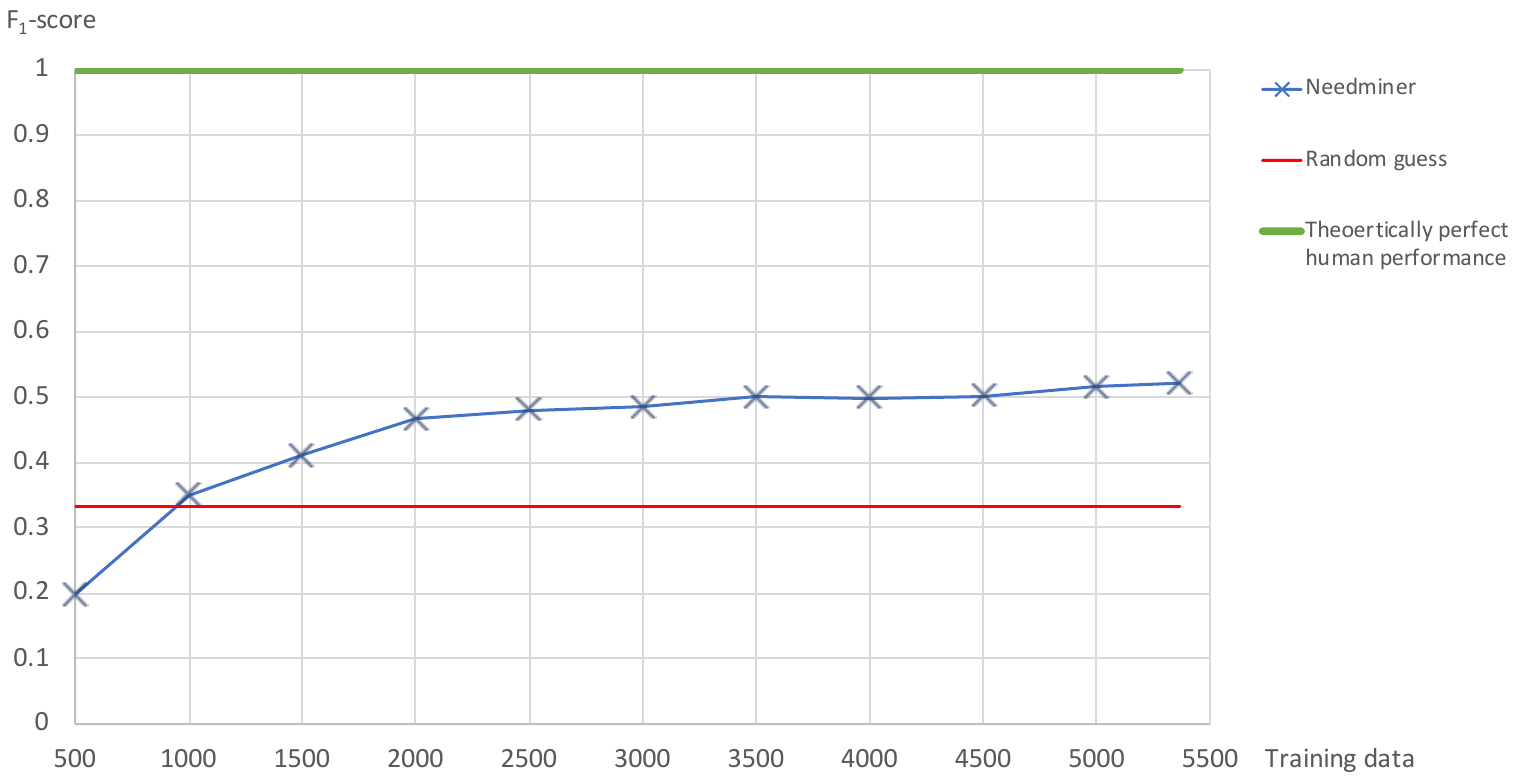}
	\caption{Learning curve of the Needminer in terms of need tweet identification}
	\label{fig:learningcurve}
\end{figure}

The criteria of \textit{expense/effort} can be divided into the stages of preparation, execution and analysis \cite{griffin1993}. In terms of preparation, Needmining requires a training phase for being set up. In particular, \Cref{fig:learningcurve} illustrates the learning curve of the Needmining artifact on the detection of need tweets in the domain of e-mobility. It reaches its maximum with the highest available amount of training data ($\sim$5.400 tweets). If we assume that additional data only provides minimal additional performance below 0.5 percentage points, we could argue that the score reaches an economically-meaningful point with $\sim$5.500 tweets as training data---meaning the cost for additional training results in highly disproportional performance gain with a reduced marginal utility \citep{kauder2015}. In our observations of the labeling process, one tweet takes $\sim$20 seconds to be labeled. This means that after roughly 30 man-hours the artifact is set up and ready to make predictions. In comparison to the task of manually looking through tweets, this means Needmining has the potential to be economically cheaper after 30 hours of training, as the artifact can predict from that point on with the identified performance. It has to be kept in mind, though, that the performance of the classifier is not comparable to the one of a human labeler, as illustrated in \Cref{fig:learningcurve}. In conclusion, the quality of the output of the artifact is not perfect and the preparation effort is noticable, while the effort for execution and analysis should be negligible.

In terms of required \textit{competencies} \citep{edvardsson2012}, the artifact does not expect any competencies from the user in advance and, thus, adds a (so far not available) method suited for non-expert users. Its \textit{suitability for innovation} \citep{cooper1987} remains to be explored and future work needs to address this aspect. Examples of need tweets show that in accordance with the labeling results, \textit{latent need detection} \citep{herstatt2007} is given in some cases, although this aspect is not generalizable. The \textit{representativeness} \citep{bryman2015} of the Needminer results remain limited. Twitter data might not be representative of an entire society and might be biased towards people who are more open to new technologies or younger people in general \citep{marshall2015}. Additionally, (obviously) only Twitter users with their specific demographics and treats are considered \citep{correa2010}. Future work needs to assess the representativeness of needs expressed on Twitter in comparison to other data sources of need elicitation. Finally, the aspect of \textit{scalability} \citep{griffin1993}---one of the initial motivations for the artifact---is given. Also in this respect, Needmining has the potential to well enhance the portfolio of existing methods.

With the demonstration of feasibility (and superiority over naive approaches) we have shown that Needmining provides a low-effort and highly scalable method for need elicitation whose use does not require expert competencies. Ultimately, the innovation potential provided by Needmining as an augmentation of existing methods needs to be verified by long-term studies in real world applications.
\section{Conclusion}
\label{sec:conclusion}

With Needmining we propose an approach to automatically identify and quantify needs from social media. We instantiate an artifact for the domain of e-mobility with German tweets as a data source to show the feasibility of the approach. The artifact is designed and implemented within three Design Science Research (DSR) cycles. The resulting software artifact demonstrates the capabilities and is already able to perform the task of automated need elicitation (with limitations).

In the first DSR cycle we analyze the \textit{feasibility} to automatically identify ``need tweets''---Twitter instances, which contain (any) customer need information, authored and posted by real users. To achieve this, we apply supervised machine learning on the basis of pre-labeled data. In multiple lab sessions, we let over 50 participants label almost $7,000$ tweets to reach a solid basis for the further model training. We then experiment with different text preprocessing, algorithm and sampling techniques. We estimate the errors of our candidate models and are able to identify feasible classification models, which reach $F_1$-scores on average of up to $0.521$ on unseen data, which is significantly better than a a random guess (+82\%) or a simple assignment of all tweets as need tweets (+56\%).  

As this classification model ``only'' predicts need tweets so far, we are still missing automated insights about the types of needs. To tackle the challenge of automatic need identification and quantification on the basis of need tweets, we employ a second cycle. We choose a supervised machine learning approach and train and evaluate eight machine learning models, one for each e-mobility need category. With these models, we are able to reach $F_1$-scores of, for instance, $0.721$ for the need category of \textit{charging infrastructure}. The improvements compared to the best performing baselines range from $+35\%$ to $+380\%$.

As we now obtain two feasible classifiers, one for predicting need tweets and one for predicting quantities of needs from need tweets, we acquire a solid basis to demonstrate the possibilities of a working Needmining software artifact in a third cycle. We suggest and implement an appropriate architecture, which is based on different web services in a service-oriented architecture. Encapsulated services handle the data collection, data storage, need tweet prediction, need prediction, gender prediction, sentiment prediction, user interface as well as the orchestration of these services. We evaluate the resulting artifact in a workshop with an industry partner. The participants of the workshop work in the field of innovation and their daily routine requires them to know about the customer needs. The majority of participants regard the artifact in its current state as helpful and can imagine to establish it within the business. The test users also point out manifold improvements, which could be implemented in future iterations of the artifact.

\subsection{Limitations \& Outlook}
The research at hand, of course, is subject to a number of limitations. As we instantiated Needmining on a specific data source (Twitter), language (German), and domain (e-mobility), each of these choices we made in the design process depicts a limitation. Other domains of interest could result in different performances. Furthermore, we only tested Needmining with German Twitter data. Other data sources might have other specificities, e.g. users using other types of language (less/more colloquial), other languages in general (English, Spanish, etc.) as well as longer texts ($>140$ characters\footnote{At the time of the data collection, Twitter only supported 140 characters per tweet.}). The regarded platforms, like Twitter, are certainly a limiting factor in terms of representativeness\footnote{E.g., these people are younger by trend \citep{correa2010}, have certain personality structures \citep{marshall2015}, and are moreover not representative for the whole society \citep{tufekci2014}}. This is a common issue in social media research and we address it partially by implementing a web service, which predicts the gender of the users to reveal more insights about demographics \citep{hirt2017b}. Nonetheless, future research needs to include even more personal information like the age, location, but also personality factors like the ''Big Five`` \cite{allport1936}. Apart from the limits of the data source, the application to e-mobility as the only domain could further restrict the insights from this research. The application to other domains could result in lower classification performances, despite new training with labeled data. Even though the current classification models achieve superior statistical performances compared to all baselines, they are far from perfect. An $F_1$-score of $1$ would denote a perfect classifier and our results show performances which are considerably less good. This might have two reasons: Either we do not utilize the full potential of natural language processing and machine learning capabilities, and/or there are some instances which simply cannot be recognized by machine learning approaches. Future research can further investigate both possibilities.  

The contribution of Needmining to innovation management and the process of product or service innovations is, so far, only ``lightweight''. We conducted only one workshop with a specific industry partner. While the results are promising, it remains to be shown how a Needmining artifact would contribute to the daily life of an innovation manager in the long run. It is still unclear, how Needmining can \textit{enhance} traditional methods of need elicitation. At the current state, while Needmining offers a lot of augmentation potential (as discussed in section 7.2 ), it is not able to \textit{replace} any existing methods, as the results from the workshop show. Another aspect regarding the practical usability is the granularity of the needs. The workshop demonstrates how different participants are interested in different levels of abstraction of the needs: Some are interested in very precise demands (e.g. ``I want a Tesla Model S in 2018''), while others are interested in more abstract or even latent needs (e.g. the need for self-expression). Apart from the abstraction level of the need, it remains to be seen what Needmining users do with the information they are presented with. In particular, it remains to be investigated if this need information ultimately can be successfully translated into innovations.

\subsection{Implications}
In contribution to the body of knowledge, we are able to demonstrate the feasibility of Needmining. This contributes to the concept of \textit{fuzzy front end} \citep{khurana1997}, where the academic discussion is still centered around possibilities to elicit customer needs in the early stages of innovation processes in a structured way \citep{alam2006, wowak2016, schweitzer2018}. As shown, Needmining could prove to be useful in this stage.

As our design is guided by justificatory knowledge from supervised machine learning, we implement a NLP- and machine-learning-based artifact and illustrate the possibility to automatically identify and quantify customer needs from Twitter data---setting the foundations for a continuous monitoring of customer needs. This can constitute a meaningful enhancement for the field of innovation management as it allows automated analytical support for the need elicitation process, which is traditionally mainly manual work. Therefore, it addresses one of the main research priorities in the area of service design---the ``utilization of big data to advance services''---as stressed by \cite{ostrom2015}. Needmining is largely based on natural language processing and machine learning. The results demonstrate that machine learning has the capability to identify complex constructs like a need, expressed by a human being. Needs can have different natures, e.g. be subtle or explicit, and be expressed in many ways, e.g., colloquial or formal. It is remarkable how well machine learning is able to capture such, even for a human being, complex situations. These insights support the work of \cite{muller2016}, who predict a rapid increase in the (cognitive) capabilities of machine-learning based artificial intelligence for the next decade. At the same time, they are a proofpoint for the potential of social media analytics \citep{stieglitz2014}.

With respect to managerial implications, Needmining could already improve certain tasks of innovation managers. Applying a fully automated analysis to a set of publicly available data can lead innovation managers to find those ``needles in the haystack'' that contain valuable need information, quantify them, and do both on an ongoing basis. These insights might help innovation managers to prioritize future endeavors---or assist them in coming up with completely new ideas.

Finally, the contribution of Needmining to innovation management in general and to customer-oriented need elicitation processes like Design Thinking in particular are topics for future research. It remains to be seen, how Needmining can contribute to the daily life of innovation managers if continuously used, where it enhances their decisions and how it generally competes with traditional need elicitation methods. This step is critical, but, if successfully shown, Needmining can be a promising enhancement for innovation management by providing analytical support for customer need elicitation.
\newpage
\bibliographystyle{apalike}
\bibliography{bibliography.bib}
\end{document}